\documentclass[preprint]{aastex}

\newcommand{\snfrac}{ {\cal F}_{\rm SN}}

\def\lta{\,\raise 0.3 ex\hbox{$ < $}\kern -0.75 em
 \lower 0.7 ex\hbox{$\sim$}\,}
\def\gta{\,\raise 0.3 ex\hbox{$ > $}\kern -0.75 em
 \lower 0.7 ex\hbox{$\sim$}\,} 
\newcommand{\sfr}{\Gamma_{\rm SF}} 
\newcommand{\mdot}{{\dot M}_c} 
\newcommand{\yield}{ \langle M[A] \rangle_\ast} 

\shorttitle{FUV Flux in Young Embedded Star Clusters}
\shortauthors{Fatuzzo \& Adams}

\newcommand{\be}{\begin{equation}}
\newcommand{\ee}{\end{equation}} 

\begin{document}

\title{Distributions of Long-Lived Radioactive Nuclei 
Provided by Star Forming Environments}  

\author{Marco Fatuzzo}
\affil{Department of Physics, Xavier University, Cincinnati, OH 45207}   

\and 

\author{Fred C. Adams}
\affil{Physics Department, University of Michigan, Ann Arbor, MI 48109}

\begin{abstract}

Radioactive nuclei play an important role in planetary evolution
by providing an internal heat source, which affects planetary
structure and helps facilitate plate tectonics. A minimum level of
nuclear activity is thought to be necessary --- but not sufficient ---
for planets to be habitable. Extending previous work that focused on
short-lived nuclei, this paper considers the delivery of long-lived
radioactive nuclei to circumstellar disks in star forming regions.
Although the long-lived nuclear species are always present, their
abundances can be enhanced through multiple mechanisms. Most stars
form in embedded cluster environments, so that disks can be enriched
directly by intercepting ejecta from supernovae within the birth
clusters. In addition, molecular clouds often provide multiple
episodes of star formation, so that nuclear abundances can accumulate
within the cloud; subsequent generations of stars can thus receive
elevated levels of radioactive nuclei through this distributed
enrichment scenario. This paper calculates the distribution of
additional enrichment for $^{40}$K, the most abundant of the long-lived radioactive nuclei. 
We find that distributed enrichment is more effective 
than direct enrichment. For the latter mechanism, 
ideal conditions lead to about 1 in 200 solar systems being directly enriched in 
$^{40}$K at the level inferred for the early solar nebula (thereby doubling the abundance). 
For distributed enrichment from adjacent clusters, about 
1 in 80 solar systems are enriched at the same level. 
Distributed enrichment over the entire molecular cloud is 
more uncertain, but can be even more effective.

\end{abstract}


\section{Introduction} 
\label{sec:intro}  

The chemical composition of circumstellar disks is important both for
their evolution and for the properties of the planets they ultimately
produce. In previous work, a great deal of attention has been given to
the short-lived radioisotopes (SLRs), such as $^{26}$Al and $^{60}$Fe
(e.g., see the results from \citealt{camtur} to \citealt{mishra}).
Meteoritic evidence indicates that the early solar system was enriched
in several species of SLRs, especially $^{26}$Al, and these nuclei
provide vital sources of both heating and ionization to the early
solar nebula and other planet-forming disks \citep{cleeves,umebayashi}. In
addition, as outlined below, long-lived radioisotopes (LLRs) can also
play an important role in the long-term evolution of planets. Whereas
SLRs decay quickly and their inventory in disks must be produced
locally, LLRs have a baseline contribution from the background nuclear
supply of the galaxy. On the other hand, LLRs are produced via
supernovae (e.g., \citealt{mathews,timmes}), which are often
associated with star forming regions. As a result, these supernovae
can enrich nearby circumstellar disks with an extra complement of LLRs
(in addition to the LLRs present in the original molecular cloud
material). The goal of this paper is to quantify the additional
enrichment of disks with LLRs due to supernovae in star forming
regions.

The chemical composition of the disk is important not only for planet
formation in general, but also for the formation of habitable planets
in particular. The most basic requirements for habitability are often
taken to be [1] the planetary mass is comparable to Earth, and [2] the
stellar insolation is comparable to that received on Earth so that the
planet can retain liquid water on its surface over geologically
interesting time scales \citep{kasting,lunine,scharf}. Moving beyond
these basic necessities, many authors have suggested that habitable
planets require additional chemical constraints \citep{kasting,gonzalez}.  
As one example, an important variable for the formation of planets is
the surface density of solid material in the disk, so that planet
formation is favored in systems with high metallicity. On the other
hand, recent results indicate that terrestrial planet formation is not
as sensitive to metallicity as the formation of larger bodies
\citep{buchave}. Moreover, habitability is thought to require a
sufficient level of geological activity \citep{efrank}, including
interior heat production and crustal recyling (e.g., active plate
tectonics). This activity is enhanced by LLRs such
as $^{40}$K, $^{235}$U, $^{238}$U, and $^{232}$Th. 
Of these, $^{40}$K is the most abundant and best understood.  In 
contrast, the actinides are produced solely via the r-process.  
While the conditions necessary for the functioning of the r-process 
are well understood, theoretical models fail to produce the full gamut 
of r-nuclides expected. As such, uncertainty remains as to both the
mechanism and astrophysical site of actinide production.
We therefore focus solely on the delivery of $^{40}$K 
to circumstellar disks in their early evolutionary phases.

Radioactive nuclei can be delivered to planet-forming disks through
two conceptually different mechanisms. In the direct enrichment
scenario, supernovae explode within the birth cluster and provide
radioactive nuclei to any circumstellar disks that are favorably
positioned at the time of detonation. This channel of enrichment is
important for SLRs; because of their short half-lives, SLRs must be
incorporated into disks on short time scales. In previous work
(\citealt{afnuke}; hereafter Paper I), we found the distributions of
SLR abundances provided to solar systems through this channel of
enrichment. Clusters can account for the abundances of SLRs inferred
for our solar system, but only $\sim10\%$ of the time; typical
enrichment levels are 10 times lower. As star formation continues,
radioactive nuclei are injected into the background molecular cloud,
where they are available to enrich the next generation of stars and
disks \citep{gounelle08,gounelle09}. This distributed enrichment
process can compete with direct enrichment for the case of SLRs. In
this paper, we explore both the direct and distributed enrichment
scenarios for the case of long-lived radioisotopes. We find that distributed 
enrichment is more effective than direct enrichment for the 
LLRs, as expected. However, a small fraction of solar systems 
(of order 1\%) can experience substantial enrichment, through 
either mechanism. As a result, a small fraction of potentially 
habitable planets are predicted to have radioactive complements 
that exceed that of Earth. Although the fraction is low, the 
total number of such planets in our Galaxy could still number 
in the billions.

This paper is organized as follows. In \S 2 we
review the nuclear yield for $^{40}$K, and obtain IMF averaged values under various scenarios. 
The resulting enrichment levels for direct injection of
nuclei into disks within a cluster environment are presented in  \S3. We
present distributions of enhancements for individual solar
systems for a distribution of clusters that mimics the observed local cluster
distribution, and also a distribution 
of clusters that extends up to $10^6$ members. We consider cases of distributed enrichment in
\S4, which presents results of a neighboring cluster enrichment 
scenario, and develops a model of radioactive
abundances over the lifetimes of molecular clouds. 
For both the direct enrichment
scenario (\S 3) and the case of enrichment by neighboring
clusters (\S 4.1), the abundances of $^{40}$K are determined by a
straightforward (but complicated) accounting calculation, and can be
directly compared; for the case of distributed enrichment over the
whole molecular cloud, however, we present only a simple model, which 
contains more uncertainties (see \S 4.2).
The paper
concludes, in \S5, with a summary of the
results and a discussion of their implications for the formation of
habitable planets.

\section{Abundance and Nuclear Yields for $^{40}$K}
\label{sec:input} 

The abundance of elements throughout the Milky Way represents a fundamental
field in astronomy that informs our understanding of stellar evolution, Galactic
chemical evolution, and planet formation.  Perhaps not surprisingly, variations in elemental 
abundances have been observed in nearby F and G stars in the Galactic thin disk
(e.g., Reddy et al. 2003), and abundances generally decrease with galactocentric 
radius.  For purposes of our investigation, we use early (4.56 Gyrs ago) solar system abundances 
as derived from chondritic meteorites and the solar photosphere 
to help guide the analysis of our results (Lodders 2010).
Those values normalized to $10^6$ Si atoms are presented in Table 1 for the most active
long-lived radioactive isotopes,  along with corresponding half-lifes $t_{1/2}$ and relative activity 
$\dot N_S \equiv \lambda N_S$ (where $\lambda = \ln[2] / t_{1/2}$).  In addition to $^{40}$K, we include
for comparison the parameters for Uranium and Thorium; note that their relative activity is much lower
than that of $^{40}$K.
 \begin{deluxetable}{cccc}
\tablecolumns{4}
\tablewidth{0pc}
\tablecaption{Table of long-lived nuclear isotopes}
\tablehead{
\colhead{Isotope}  
& \colhead{$t_{1/2}$ (Gyrs)} &
\colhead{$N_s$}  &   \colhead{$\dot N_s$ (Gyrs$^{-1})$}}
\startdata
$^{40}$K  & 1.25 & $6$ & $3.3$  \\
$^{232}$Th  & 14.1 & $0.0440$ & $0.0022$ \\
$^{238}$U  & 4.47 & $0.0180$ & $0.0028$ \\
$^{235}$U  & 0.704 & $0.0058$ & $0.0057$\\
\enddata
\end{deluxetable}

The production of $^{40}$K occurs primarily in the supernova events that mark 
the death of massive stars.  Detailed calculations of nucleosynthesis yields of this isotope as a function
of progenitor star mass appear in the seminal paper by Woosley \& Weaver 
(1995 - hereafter WW95) and the follow up paper by Rauscher et al. 
(2002 - hereafter R02), although the latter
does so for a narrower 15 -- 25 $M_\sun$ mass range.  We therefore update the $^{40}$K yields 
presented in WW95 (which spans 8 -- 40 $M_\sun$) with those presented in R02 
over the 15 - 25 $M_\sun$ range.  
In order to get yields for any specified mass above the
supernova threshold of 8 $M_\sun$, 
we adopt a linear interpolation scheme in log-log space,
and extrapolate from the endpoint values down to $8 M_\sun$ and
up to an assumed maximum stellar mass of $120 M_\sun$.  
Yield rates and interpolated/extrapolated values are presented in Figure 1.
Although our analysis incorporates yields calculated for progenitor stars 
with solar metallicity, stellar evolution could affect our results.  
Toward that end, we note that WW95 calculate yields  for progenitor stars for different
metallicities $Z/Z_\sun$. We present their full results for $^{40}$K in Figure 2. 

\begin{figure}
\figurenum{1}
{\centerline{\epsscale{0.60} \plotone{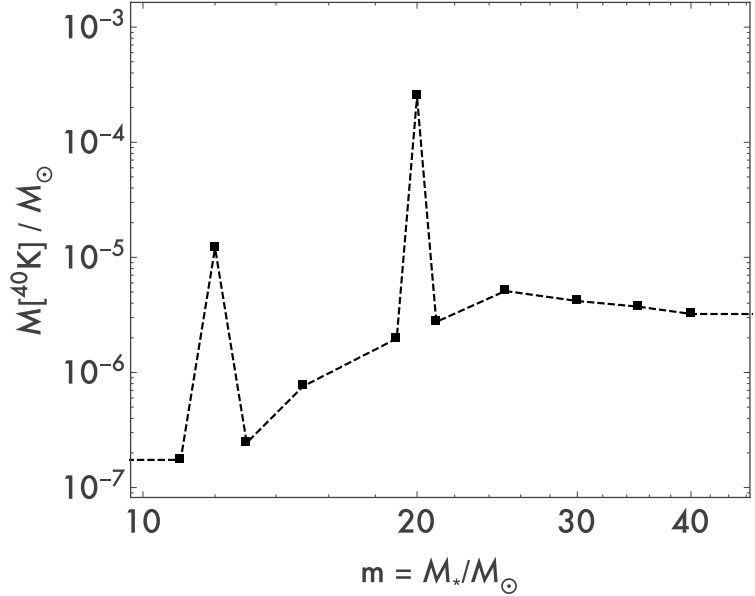} }}
\figcaption{Calculated radioactive yield $M[^{40} {\rm K}]$ as a function of progenitor mass  $M_*$
based on the results from WW95 and R02.  
The dashed line represents the adopted interpolation / extrapolation scheme as outlined in the text of the paper.  } 
\end{figure}

\begin{figure}
\figurenum{2}
{\centerline{\epsscale{0.60} \plotone{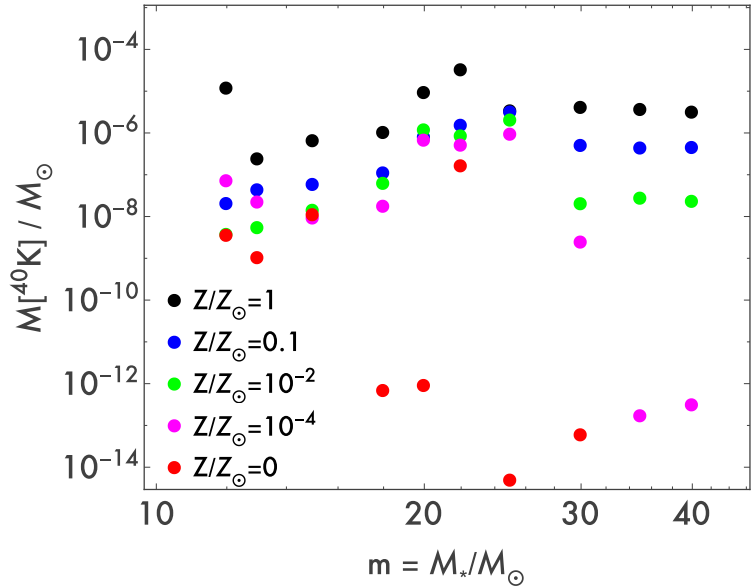} }}
\figcaption{Calculated radioactive yield  $M[^{40} {\rm K}]$ as a function of progenitor mass for several metallicities 
(Woosley and Weaver 1995). } 
\end{figure}

We next consider the yield per star expected from a population of stars born with a specified initial mass function (IMF).  
Following Adams et al. (2014), we parametrize the high mass portion of the stellar IMF that leads to 
supernova events through the power-law relation
\be
{dN_*\over dm} = \snfrac \,{\gamma\over 8}\left({m\over 8}\right)^{-(\gamma+1)}\,,
\ee
where $\snfrac$ is the fraction of the (initial) stellar population with mass greater
that the minimum mass $m_{min}=8$ (in solar units) required for a star to end its life with a 
supernova explosion, $m$ is the stellar mass in solar units, and $\gamma$
is an index value.  From observations, we expect  ${\sc F_{SN}}\approx 0.005$ (e.g., 
Adams \& Fatuzzo 1996; Kroupa 2001; Chabrier 2003) and $\gamma\approx 1.35$
(e.g., Salpeter 1955).  But given the observed scatter in the index $\gamma$ along with 
its observational uncertainty, we consider a range of $1\le \gamma \le 2$ in our analysis.
Note that the relation given in Equation (1) is normalized so that
\be
\int_8^\infty {dN_*\over dm} \,dm = \snfrac\;,
\ee
and thus requires a correction factor of
$f_C = [1-(8/m_\infty)^\gamma]$ for an IMF with an upper mass limit $m_\infty$.
This correction factor is $f_C = 0.983$ for a fiducial value of $m_\infty = 120$ and
an index $\gamma = 1.5$, and
since a $\sim 2$ percent correction is much smaller than the other uncertainties 
in the problem, we will generally ignore it. 

The yield per star weighted by a stellar IMF with index $\gamma$ is 
then found through the integral expression
\be
\langle M[^{40} {\rm K}]\rangle_* \equiv \int_8^\infty M[^{40} {\rm K}; m] {dN_*\over dm} \,dm\,,
\ee
where $M[^{40} {\rm K}; m]$ is the yield of $^{40}$K as a function of progenitor mass $m$
(as illustrated by the dashed line in Figure 1 for our adopted interpolation/extrapolation scheme). 
The yield per star for $^{40}$K is shown in Figure 3 as a function 
of $\gamma$.  Note that the yield per supernova weighted by a stellar IMF with index $\gamma$
is given by the relation
\be
\langle M[^{40} {\rm K}]\rangle_{SN} = {\langle M[^{40} {\rm K}]\rangle_* \over {\cal F}_{SN}}\;.
\ee
\begin{figure}
\figurenum{3}
{\centerline{\epsscale{0.60} \plotone{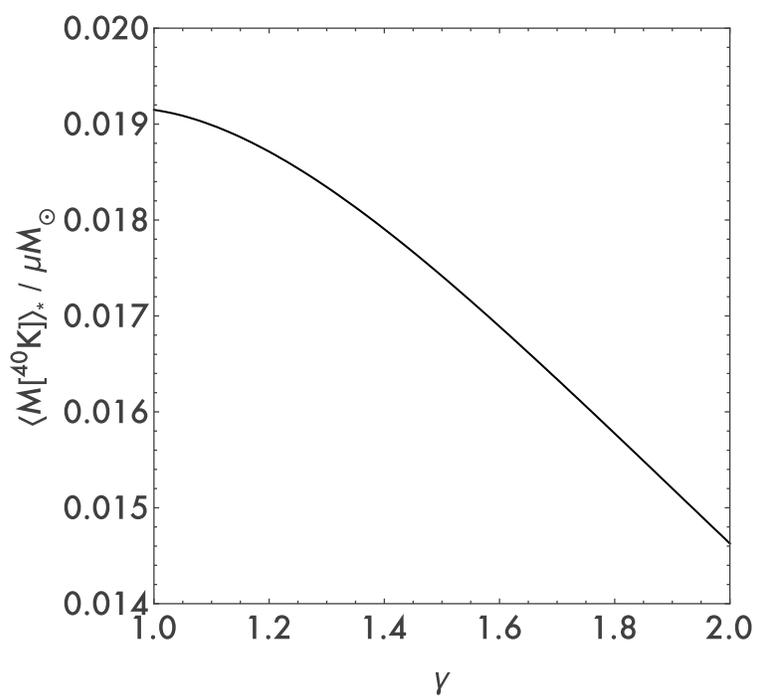} }}
\figcaption{Radioactive yield per star for $^{40}$K versus index $\gamma$ of the stellar IMF. The
yields, which are given in units of $\mu M_\sun$, are proportional to the fraction ${\cal F}_{SN}$ of stars
above the supernova mass threshold, taken here to be 0.005.}
\end{figure}

The ``background" value for $^{40}$K as defined by its number ratio to hydrogen
$(N_{^{40} {\rm K}}/N_H)_{BG}$ can be estimated by assuming a
uniform mixing of SN ejecta with the interstellar medium.  Adopting a constant fiducial rate ${\cal R}_{SN} = 
0.01$ yr$^{-1}$ of supernova events, constant yields obtained using an IMF index of $\gamma = 1.5$
and solar metallicity, and using a  fiducial value of $M_H
 = 5.5\times 10^9 M_\sun$ for the hydrogen mass in the ISM, 
the steady-state
abundance $M[^{40} {\rm K}]_0$ is easily found by balancing the
injection and decay rates
\be
{\cal R}_{SN} \langle M[^{40} {\rm K}] \rangle_{SN} = \lambda M[^{40} {\rm K}]_0\;,
\ee
which then yields a value of $(N_{^{40} {\rm K}} / N_H)_{BG} = 2.8\times 10^{-10}$.  In contrast,
the early solar system value is $(N_{^{40} {\rm K}} / N_H)_{SS} = 2.3\times 10^{-10}$ (Lodders 2010).
We note that our idealized analysis ignores evolution effects both in the rate of supernovae and the yields (which
are strongly dependent on metallicity $Z$ as shown in Figure 2).  In addition, a significant fraction of the
isotope mass is likely locked up in stars and stellar remnants.  In all, our estimate is therefore likely an upper limit
to the actual value.

\section{Cluster Self--Enrichment Distributions}
We consider first a scenario where enrichment occurs within a
cluster due to its own members.  In this case, only the most massive stars would evolve
on short enough timescales to affect disk evolution.  Specifically, circumstellar disks are expected to retain 
their gas for $\sim 3 - 10$ Myrs, and only $M_* \gtrsim 16 M_\sun$
stars evolve all the way to core-collapse on comparable timescales.  As a way of quantifying how 
stellar evolution affects the yield of long-lived radioactive isotopes within a single cluster,  we calculate 
nuclear yields per star for different values of $m_{min}$ of a parent IMF with spectral index $\gamma = 1.5$. 
As shown in Figure 4, the yields (per star) of $^{40}$K, which decreases steadily with increasing minimum mass, is fairly 
sensitive to $m_{min}$, and hence to cluster and
stellar disk evolution within a cluster environment.   To further illustrate this point, we plot
in Figure 5 the nuclear yield (per star) of $^{40}$K as a function of cluster age $\tau$ assuming that only stars that have evolved
to core-collapse are included in the yield - that is, we match the cluster age $\tau$ to a corresponding mass 
$m_{min}$ by matching the evolutionary time of such a star to the cluster age, invoking the simple scaling law
\be
\tau ({\rm Myrs}) = 3 + {1200\over m^{1.85}}\;,
\ee
where 
this scaling law is consistent with detailed stellar 
evolution models (e.g., WW95, R02, and others) for stars 
massive enough to be supernova progenitors. If stellar disks 
lose the majority of their gas within 3 Myr, no enrichment can occur through the evolution and subsequent supernova 
explosion of a massive cluster member.  On the other hand, disk survival times in excess of $\sim 8$ Myrs allow for the
possibility of significant enrichment, owing to the fact that $^{40}$K has a peak yield at
$\sim 20 M_\sun$ (see Figure 1).

\begin{figure}
\figurenum{4}
{\centerline{\epsscale{0.60} \plotone{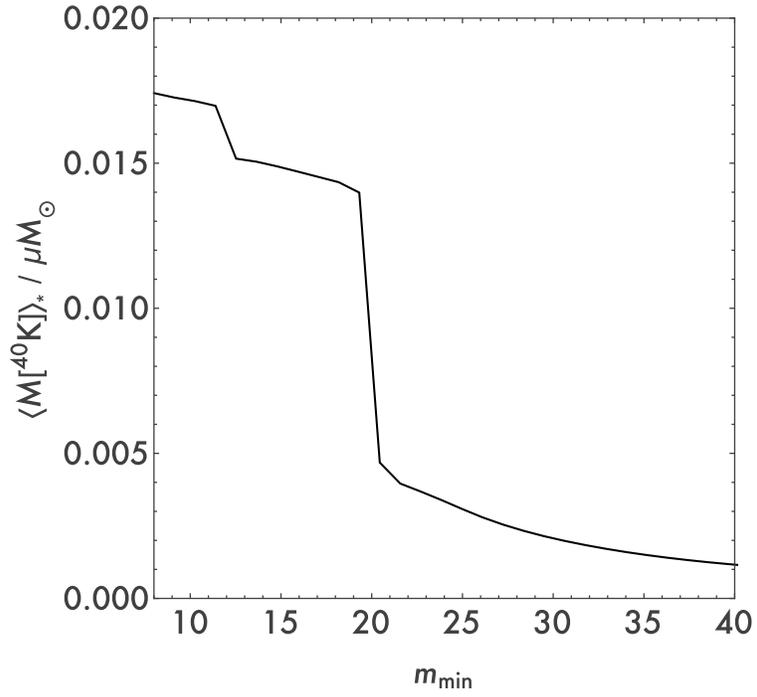} }}
\figcaption{Radioactive yield per star for $^{40}$K versus  minimum mass
of progenitor star included in the distribution for a stellar IMF with index $\gamma = 1.5$. The
yields are given in units of $\mu M_\sun$, and are proportional to the fraction ${\cal F}_{SN}$ of stars
above the supernova mass threshold, taken here to be 0.005.}
\end{figure}

\begin{figure}
\figurenum{5}
{\centerline{\epsscale{0.60} \plotone{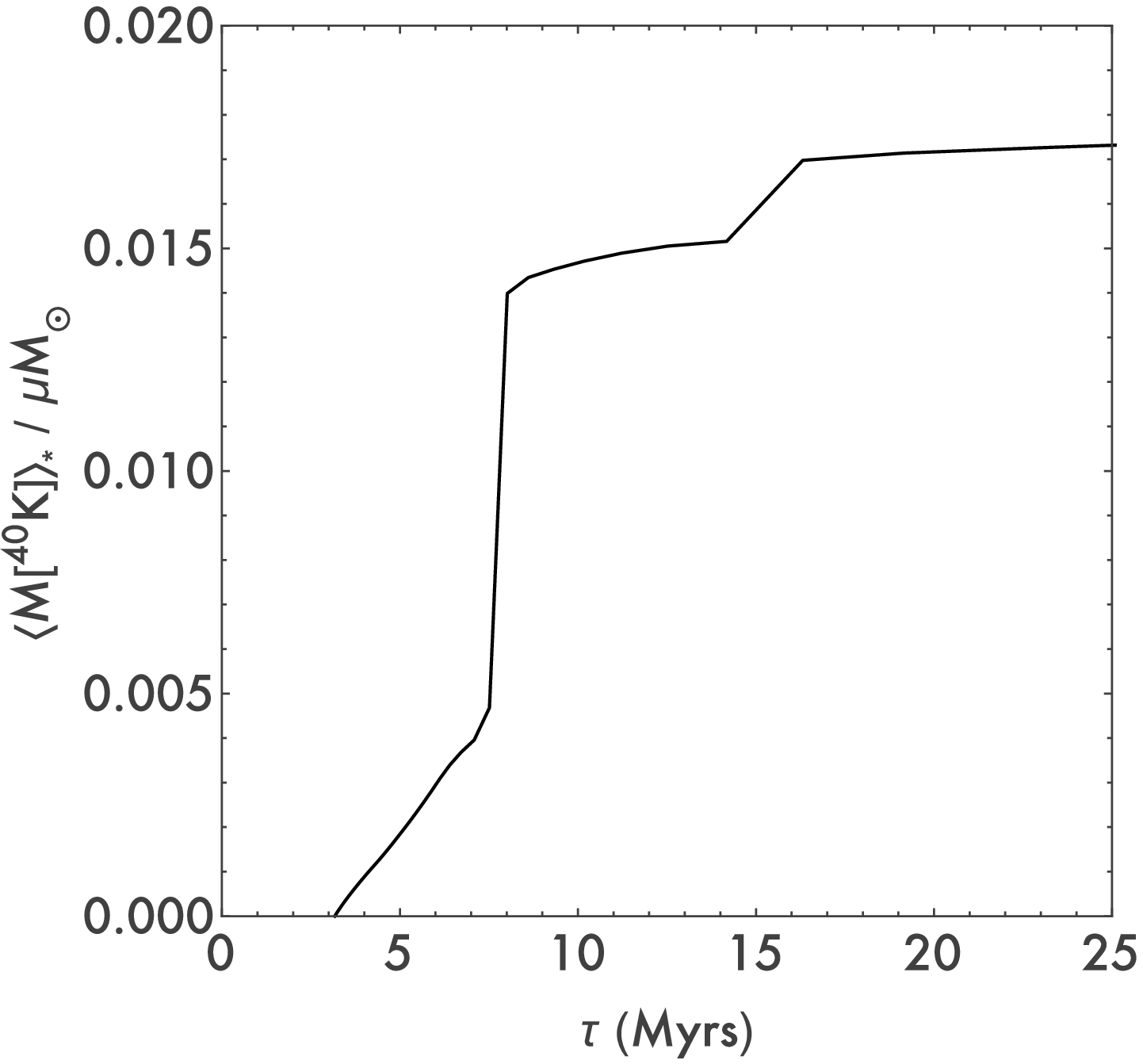} }}
\figcaption{Radioactive yield per star for $^{40}$K versus cluster age (in Myrs).
For a given cluster age,  only those stars that have evolved enough to explode as supernovae are
included in the integral over the stellar mass distribution.  The
yields are given in units of $\mu M_\sun$ and the index of the stellar IMF is $\gamma = 1.5$.
Yields are proportional to the fraction ${\cal F}_{SN}$ of stars 
above the supernova mass threshold, taken here to be 0.005.}
\end{figure}

For the sake of definiteness, we use a minimum
progenitor mass of 16 $M_\odot$ and an IMF index of $\gamma = 1.5$. This mass scale
corresponds to a main-sequence lifetime of 10.1 Myr (see equation [6]),
so that we are implicitly assuming that either circumstellar disks
retain their mass over this time, or that they form after the
progenitors. In this scenario, the fraction of stars that can 
enrich disks while they remain intact is
\be
{\cal F_{\rm c}} = \snfrac \,{1.5\over 8} \int_{16}^\infty \left({m\over 8}\right)^{-2.5}\,dm = 0.0017\,,
\ee

Studies of clusters out to 2 kpc (Lada \& Lada 2003) and out to 1 kpc
(Porras et al. 2003) indicate that in the solar neighborhood, the
number of stars born in clusters with $N$ members is (almost) evenly
distributed logarithmically over the range $N \approx$ 30 to 2000,
with half of all stars belonging to clusters with $N \la 300$.  Clearly, stellar disks
in small clusters have a small probability of being enriched as a result of a SN event
within the cluster,  whereas their couterparts
in large clusters will likely be enriched from several SN events.  

We now calculate enrichment distributions for stellar disk systems in the solar neighborhood.
A typical stellar disk in a cluster will intercept a fraction
\be
f(r) = {\pi R_d^2\over 4 \pi r^2}\,\cos\theta\,,
\ee
of the radioactive yield produced by 
the entire cluster, where $R_d$ is the disk radius, and $r$ is the distance from the
stellar disk to the cluster center (where the high-mass stars, and hence the supernova
ejecta, originate).  The factor of $\cos\theta$ takes into account the fact that the
disk is not, in general, facing the supernova blast wave.  For simplicity, we replace
$\cos\theta$ with its mean value of $1/2$.  The radius $r$ must be larger than the 
radius for which the disk (with radius $R_d$) is stripped due to the blast; for 
a disk of radius $R_d = 30$ AU, and for typical supernova energies, this minimum
radial distance is $r_{min}\approx 0.1$ pc (Chevalier 2000; Ouellette et al. 2007; Adams 2010).
The largest capture fraction expected is therefore $f_{max} \approx 2.6\times 10^{-7}$.
In turn, the maximum mass of $^{40}$K that can be captured by a disk for a single
SN event is found to be
$65$ p$M_\sun$ (p = $10^{-12}$), although that requires an ideal progenitor mass of $M_* = 20 M_\sun$.
More likely capture values for stellar disks located $\sim 0.1$ pc from a single progenitor would be
$\sim 0.1$ p$M_\sun$, so distributions for local (Lada \& Lada) clusters would be expected to range
between $\sim 10^{-3}$ p$M_\sun$ and $\sim 0.1$ p$M_\sun$, with a high yield ``wing'' that
extends up to $\sim 10^2$ p$M_\sun$.  
 
To generate the distributions of captured mass,  we first pick the cluster size that our ``target" disk populates through a 
random sampling of a Lada \& Lada cluster distribution, where the sampling 
uses a probability function that assigns the cluster size based on the probability of a star
being in such a cluster (as opposed to the probability of a cluster having a given size).  We then select the masses of each
star within the cluster through a random sampling of the IMF, adopting a value of $\gamma = 1.5$
for our analysis.  The cumulative yield of each isotope that results from the evolution of
$ \ge 16 M_\sun$ stars is then calculated, and assumed to originate from the cluster center.  
The location of a disk system is then randomly picked on the assumption that stars are distributed 
throughout the cluster in accordance to an average gas density profile $\rho_* \propto 1/r^\beta$,  
where observations indicate that $\beta$ ranges between 1 and 2.  We select a value of $\beta = 2$
in order to maximize the captured mass (since more stars are located near the cluster center),
leading to the relation $dm \propto r^2 \rho_* dr \propto dr$.  
The cumulative probability that a star/disk system in a cluster with $N$ members
is located at radius $r$ is then given by
\be
P(r) = \left({r\over R_c}\right)\,,
\ee
where the outer boundary $R_c$
is set through the empirically  determined relation between cluster radius and number of stars 
\be 
R_c (N) = 1\,{\rm pc}\,\left({N \over 300}\right)^{1/2} \,,
\label{rofn}
\ee 
(see Figure 2 of Adams et al. 2006, which
uses the data from Carpenter 2000 and Lada \& Lada 2003).  
If the radius is smaller than $0.1$ pc, the disk is assumed
to not survive, and the result is not included in the distribution.  Otherwise, the total yield from the
progenitor stars is multiplied by the capture fraction $f$.  The process is then repeated
$100,000$ times, with the resulting distribution presented in Figure 6.
We then repeat the same process for a Lada \& Lada type cluster distribution that extends up 
to $N = 10^6$, but adopt the more realistic large-cluster scaling
\be 
R_c (N) = 1\,{\rm pc}\,\left({N \over 300}\right)^{1/3} \,.
\label{rofn}
\ee 
The resulting distribution  is presented in Figure 7.   We note that disks within the 0.1 pc
disruption limit or for which no enrichment occurred are not represented in the histograms shown in 
Figures 6 and 7.  For completeness, we present the percentages of stellar disk systems 
enriched by $N_{SN}$ supernova events in Table 2 for both a local Lada \& Lada distribution
(LL) and for a Lada \& Lada type cluster distribution that extends up 
to $N = 10^6$ (LL6).  Only 43\% of disk systems in a distribution of local (LL) clusters are
expected to be enriched by at least one SN event, but that value increases to 77\% 
for the extended (LL6) distribution
(though in both cases a fraction of disks are destroyed because they are within the 0.1 pc limit). 
Even more dramatic, the mean value of SN explosions that
enrich a disk system increases from 0.78 for the LL cluster distribution to 170 for the LL6 cluster distribution
(note that the entries in Table 2 do not add up to 100 percent for 
the LL6 cluster distribution as many clusters have even larger numbers 
of supernovae).

\begin{deluxetable}{ccc}
\tablecolumns{3}
\tablewidth{0pc}
\tablecaption{Percentage of disks enriched by $N_{SN}$ events for local (LL) and extended (LL6) cluster
distributions.}
\tablehead{
\colhead{$N_{SN}$} 
& \colhead{\% (LL)}& \colhead{\% (LL6)} }
\startdata
0 & 57&23 \\
1 &23 &9.3\\
2 & 9.8& 4.8\\
3 & 5.1 &3.5 \\
4 & 2.7 &2.3\\
5 & 1.2 &2.0\\
6 & 0.5 &1.6\\
7 & 0.2 &1.3\\
\enddata
\end{deluxetable}

\begin{figure}
\figurenum{6}
{\centerline{\epsscale{0.80} \plotone{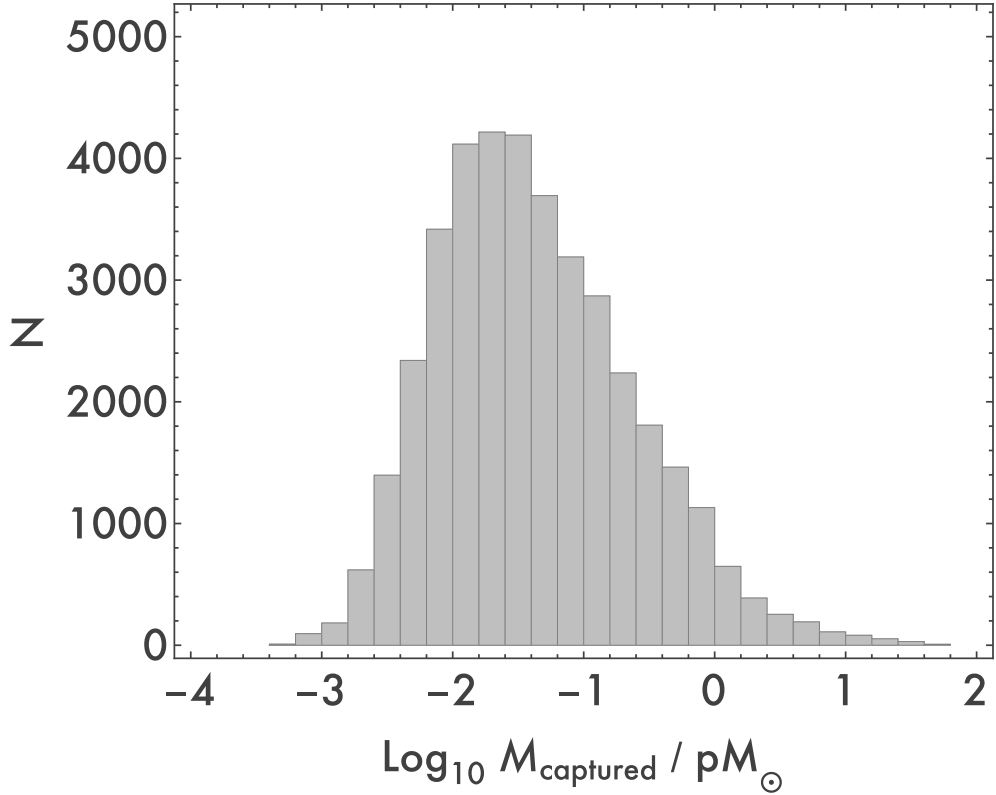} }}
\figcaption{Histogram of the $^{40}$K mass capture distribution for local clusters.  Disk systems
that were either distrupted because they were within $0.1$ pc of the cluster center, or were
not enriched due to a lack of SN events, are excluded from this distribution.  }
\end{figure}

\begin{figure}
\figurenum{7}
{\centerline{\epsscale{0.80} \plotone{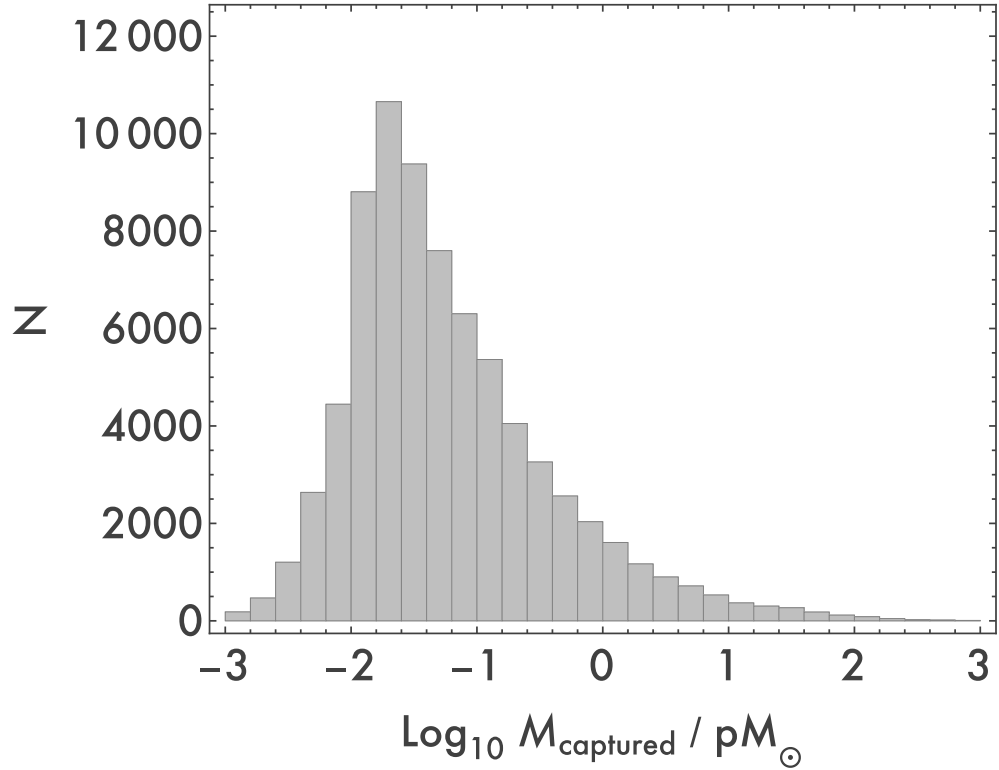} }}
\figcaption{Histogram of the $^{40}$K mass capture distribution for clusters that have a Lada and Lada
distribution that extends to $N = 10^6$ stars. Disk systems
that were either distrupted because they were within $0.1$ pc of the cluster center, or were
not enriched due to a lack of SN events, are excluded from this distribution. }
\end{figure}

To put our results in context, we estimate the mass of $^{40}$K expected in a 30 AU disk
given measured solar system abundances.  Observations indicate that a 30 AU radius prototellar
disk has a total mass of $\sim 0.02 M_\sun$ (e.g., Andrews et al. 2010), and a mass of hydrogen of
$\approx 0.015 M_\sun$.  Using the 
results from \S 2, we then find the corresponding expected mass of $M[^{40}$K$]_{SS} \approx 140$ p$M_\sun$. 
The fraction of disks expected to be enriched at
a level greater than $ M_{SS}, 0.5 M_{SS}, 0.1 M_{SS}$ and $0.01 M_{SS}$ under each scenario
is given in Table 3, where LL denotes a Lada \& Lada cluster distribution, and LL6 denotes the
Lada \& Lada distribution extended up to $N = 10^6$ stars.  Note that these fractions are based on all systems sampled, including those that were not enriched or were destroyed (and are therefore not
represented in the histograms of Figures 6 -- 7).  
These results indicate that the 
fraction of systems that are enriched with $^{40}$K yields 
at the level of our early solar system is about 0.005, or about 
1 in 200. The corresponding fraction for enrichment at half 
(one tenth) of the solar system abundance is 0.009 (0.03). Although the fraction is low, the total 
number of systems in the Galaxy is large, with a corresponding 
large number of potentially habitable solar systems. If the 
prospects for habitability are enhanced by greater abundances 
of radioactive nuclei, then up to $\sim1$ billion planets in 
the Galaxy could be enriched sufficiently to satisty this criteria.

\begin{deluxetable}{lccc}
\tablecolumns{4}
\tablewidth{0pc}
\tablecaption{Fraction of disk enrichment by thresholds.}
\tablehead{
\colhead{Case} 
& \colhead{ LL }& \colhead{LL6}& \colhead{NN} }
\startdata
$M> M[^{40}$K$]_{SS}$&0& $5.2\times 10^{-3}$& $1.3\times 10^{-2}$\\
$M> 0.5 M[^{40}$K$]_{SS}$&0&$9.2\times 10^{-3}$ & $2.7\times 10^{-2}$\\
$M> 0.1 M[^{40}$K$]_{SS}$& $1.1\times 10^{-3} $&$3.0\times 10^{-2}$& 0.15\\
$M> 0.01 M[^{40}$K$]_{SS}$& $1.3\times 10^{-2}$&$0.12$& 0.52\\
\enddata
\end{deluxetable}

\section{Distributed Enrichment Scenarios}
This section considers the case of distributed enrichment 
of radioactive nuclei, where the delivery of LLRs takes place 
over longer distances (and longer time scales) than direct 
enrichment within the cluster. Note that such enrichment 
can be considered over a range of size and time scales, and 
that the problem is not as well-defined as in the case of 
direct enrichment. Here we consider two cases: In the first 
scenario (\S 4.1), the supernovae from a given cluster 
can enrich the protostellar cores of a nearby cluster. 
In the second case (\S 4.2), we consider the entire 
molecular cloud as a dynamical system and consider the 
enrichment of LLRs over its lifetime. This latter scenario is 
more uncertain, but allows for greater radioactive enrichment.

\subsection{Neighboring Cluster Enrichment Scenario} 

In this section, we consider how much enrichment can occur in dense molecular 
cores whose parent cluster neighbors another cluster that evolved at an earlier time.  
To keep the analysis as simple as possible, we assume a fiducial core radius of
$R_{core} = 0.1$ pc and core mass of $M_{core} = 10 M_\sun$.  We also assume that all stars
in the neighbor cluster with a mass $\ge 8 M_\sun$ evolve to their SN state before 
the cores in the parent cluster undergo collapse, and that the cores capture all of the 
radioactive isotopes that 
impinge upon them from the center of their neighboring cluster.  

Ensuing distributions of capture mass are calculated by first selecting the neighbor cluster
size (in terms of membership $N$) 
by sampling the local Lada \& Lada cluster distribution (as was done in \S 3), and then sampling the IMF 
($\gamma = 1.5$) to determine the mass of each star. 
The total mass yield
ejected by stars massive enough to yield SN events is subsequently calculated using the yields as shown 
in Figure 1.  The local Lada \& Lada cluster distribution is then sampled again
to determine the size of the parent cluster, and the radii of the neighbor cluster ($R_{nc}$)
and parent cluster ($R_{pc}$) are set via the same scaling -- as given by Equation (10) -- that
was used to set the cluster radius in \S 3 for the Lada and Lada cluster distribution.
A molecular core is then placed at random in the
parent cluster on the assumption that stars are distributed 
throughout the cluster in accordance to an average gas density profile $\rho_* \propto 1/r$.  We note
that setting $\beta = 1$ in the density profile (see discussion prior to Equation [9]) leads to a cumulative probability
\be
P(r) = \left({r\over R_{pc}}\right)^2\,,
\ee
that a core is located at radius $r$ from the parent cluster center.  As such, adopting this 
density profile
distributes the highest fraction of cores near the cluster
edge, and therefore maximizes the number of cores that have near maximum values of 
captured mass.
The total mass captured by a core is then obtained by assuming 
that it is able to fully capture a fraction 
\be
f = {\pi R_{core} ^2 \over 4 \pi d^2}\,,
\ee
of the radioactive material ejected from the neigboring cluster, where
the disance between the core and the neighbor cluster center
\be
d =\sqrt{(R_{nc}+R_{pc}+R_{pc}\cos\theta_c)^2+(R_{pc}\sin\theta_c)^2}\,,
\ee
is set through a random selection of the position angle $\theta_c$.
Results of the mass enrichment for the entire 10$M_\sun$ cores 
are shown in Figure 8 (in contrast, Figures 6 and 7 show mass enrichment for a 
30 AU protostellar disk).  For this scenario, 68\% of cores sampled were enriched by
at least one neighboring SN event -- higher than the percentage of disk systems
enriched for the local Lada and Lada cluster distribution because all stars
with mass $\ge 8 M_\sun$ are assumed to lead to enrichment (as opposed to the
$16 M_\sun$ limit assumed in \S 3).  Cores for which no enrichment occurred 
are not represented in the histogram shown in 
Figure 8.

As a comparison, we estimate the mass of $^{40}$K expected in a 10 $M_\sun$ core
given measured solar system abundances. Using the 
isotope ratios from \S 2, we find the corresponding expected mass of $\approx 6.8\times 10^4$ p$M_\sun$.
Enhancement values are given in column NN of Table 3, where as with the LL and LL6 columns, fractions are based on all systems sampled, including those that were not enriched (and therefore not included
in the histogram shown in Figure 8).  Roughly 1\% of solar systems are enriched with $^{40}$K with 
radioactive yields comparable to those found in the early 
solar nebula. As discussed earlier, this fraction corresponds 
to billions of planets in a galaxy the size of our Milky Way.

\begin{figure}
\figurenum{8}
{\centerline{\epsscale{0.80} \plotone{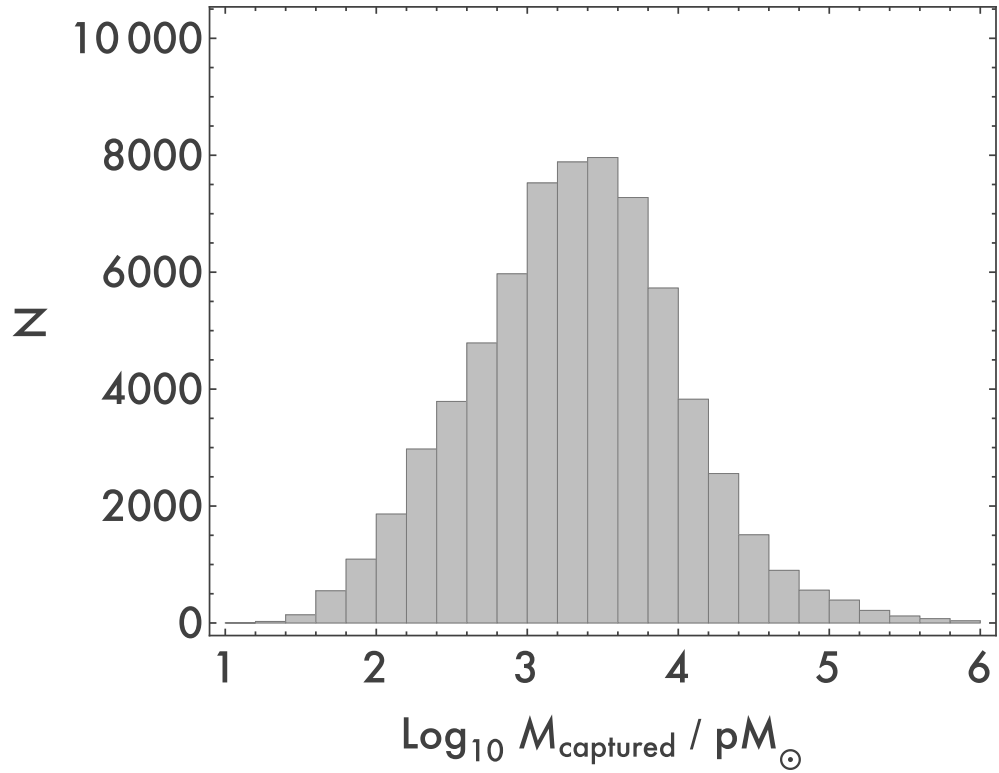} }}
\figcaption{Histogram of $^{40}$K mass capture distribution for the neighboring cluster scenario.
These abundances 
represent the total mass captured by the core; only a 
fraction of this mass will be delivered to the nebular 
disk formed by its subsequent collapse.
Molecular cores
that were not enriched due to a lack of SN events in  the neighboring cluster
are excluded from this distribution.  }
\end{figure}

\subsection{Distributed Enrichment Scenario for entire Molecular Clouds} 
\label{sec:distribute} 

In this scenario we consider a molecular cloud as a star forming 
system and study the abundance of LLRs as a function of time.  Let
$M_{\rm c}$ denote the mass of the molecular cloud and let $M_A$ denote the
total mass contained in a given isotope of interest (e.g., $^{40}$K). 

The time evolution of the entire cloud is given by the equation 
\be
{dM_{\rm c} \over dt} = - \sfr - \mdot \,,
\label{clouddif} 
\ee
where $\sfr$ is the star formation rate (in mass per unit time) and
$\mdot$ is an additional mass loss term. Star formation is generally
an inefficient process, such that only a fraction $\epsilon$ of the
cloud mass is converted into stars over a free-fall time $\tau$. In
general, we can write the star formation rate in the form
\be
\sfr = {\epsilon \over \tau} M_{\rm c} \, . 
\label{sfr} 
\ee
In the absence of the additional mass loss term $\mdot$, the molecular
cloud mass would decay exponentially with decay time scale $T =
\tau/\epsilon$. Since the free-fall time $\tau \sim 1$ Myr, and the
efficiency is low, $\epsilon$ = 0.01 -- 0.05, the decay time $T$ = 20
-- 100 Myr, which is roughly comparable to the expected cloud
lifetimes. Without additional mass loss, the cloud would still retain
1/e of its original mass at the time when it should be destroyed. The
additional term accounts for mass loss due to the disruptive effects
of stellar winds and supernova explosions, i.e., feedback processes
that act to dissipate the cloud. In this simple model, we parameterize 
the magnitude of the additional mass loss term by writing it in the form
\be
\mdot = \gamma {\epsilon M_{\rm c\,0} \over \tau} = 
 \gamma {M_{\rm c\,0} \over T} \,,
\label{mdot} 
\ee
where $M_{\rm c\,0}$ is the initial mass of the cloud. With the mass loss 
terms specified through equations (\ref{sfr}) and (\ref{mdot}), the 
time evolution of the cloud mass can be determined, 
\be
M_{\rm c}(t) = M_{\rm c\,0} \left\{ (1 + \gamma) \exp[-t/T] - \gamma \right\}\,.
\ee
Note that the cloud mass reaches zero at a time $t_f$ given by 
\be
t_f = T \log \left[ {1 + \gamma \over \gamma} \right] \,. 
\label{lifetime} 
\ee
The time $t_f$ thus represents the total lifetime of the cloud. 
We can use observations to specify the cloud lifetime $t_f$ and use 
equation (\ref{lifetime}) to determine the parameter $\gamma$, i.e., 
\be
\gamma = {1 \over \exp[t_f/T] - 1} \approx {1 \over {\rm e} - 1} 
\approx 0.58\,. 
\ee
The approximate values result from using a typical free-fall time 
$\tau$ = 1 Myr, a star formation efficiency $\epsilon$ = 0.025, and 
a cloud lifetime of $t_f$ = 40 Myr, so that $t_f/T$ = 1. 

The time evolution of the mass in a given radioisotope is then 
given by the equation 
\be
{d M_A \over dt} = \sfr { \yield \over \langle m \rangle} - 
{M_A \over M_{\rm c}} \left[ \sfr + \gamma {M_{\rm c\,0} \over T} \right] 
- {\log 2 \over t_{1/2}} M_A \,, 
\label{isodif} 
\ee
where $\yield$ is the mass of the isotope produced per star (averaged 
over the stellar IMF), and
\be
\langle m \rangle = \int {dN_*\over dm} m \,dm\,,
\ee
is the average stellar mass for a given IMF.
For the isotopes of interest, the half-lives 
are of order 1 -- 10 Gyr, whereas the cloud lifetimes are of order 0.1 Gyr, 
so we can ignore the third term in equation (\ref{isodif}). The solution 
can be written in the form 
\be
M_A(t) = \left[ M_{A0} + { \yield \over \langle m \rangle} 
M_{\rm c\,0} {t \over T} \right] 
\left\{ (1+\gamma) \exp[-t/T] - \gamma \right\} \,.
\ee

The mass fraction $F_A(t)$ of the isotope $A$ is thus given by 
\be
F_A(t) = {M_A(t) \over M_{\rm c}(t)} = 
\left[ {M_{A0} \over M_{\rm c\,0}} + 
{ \yield \over \langle m \rangle} {t \over T} \right] \,.
\ee
The mass fraction is thus a steadily increasing function of time. 
Note that the quantity $\yield/\langle{m}\rangle$ is essentially the
mass fraction (of isotope $A$) produced by the aggegrate of supernovae
in the cloud, whereas the quantity $M_{A0}/M_{C0}$ is the starting
mass fraction. We expect this second fraction to be smaller than the 
first. Moreover, the time $t/T$ is of order unity near the end of 
the cloud's lifetime, so that the mass fraction increases toward 
the benchmark value 
\be
F_A(t) \to {\yield \over \langle m \rangle} \,. 
\ee
For the case of $^{40}$K, for example, this asymptotic 
mass fraction is about $3.8 \times 10^{-8}$, which is about 
5.6 times larger than early solar system abundance of the isotope 
(with a mass fraction of about $6.8 \times 10^{-9}$ based on the Lodders 2010
values, as presented in Table 1). As a 
result, distributed enrichment over the course of a cloud's 
lifetime can -- in principle -- produce significant enrichment
of long-lived radioactive nuclei.

The enrichment levels discussed above are subject to a number of
uncertainties. This treatment of the problem implicitly assumes that
all of the LLRs produced by supernovae remain in the molecular cloud.
In practice, however, some fraction will escape. In addition, the mass
fraction only approaches its asymptotic value near the end of the
cloud's lifetime, i.e., when it retains only a small fraction of 
its original mass.  As a result, only the last generation of star
formation within the cloud would be exposed to such high levels of
radioactive nuclei. Since the mass fraction $F_A(t)$ is a linear 
function of time, the median enrichment level (in the absence of 
losses) is about half the asymptotic value, or about 3 times the
cosmic abundance. Finally, we note that molecular clouds are complex
and that supernova explosions are not uniformly distributed within 
the cloud. As a result, radioactive enrichment will not take place 
in a homogenous fashion.

\section{Conclusion} 
\label{sec:conclude} 

This paper has considered the possible enrichment of circumstellar
disks by long-lived radioactive nuclei, which are produced by
supernovae in star forming regions.These LLRs are important
components of the terrestrial planets that form within these disks.
They provide a significant internal heat source that affects the
internal structure of the planets, and helps to drive plate tectonics
and related geophysical processes.  This paper focuses of the isotope
$^{40}$K, because it is the most abundant and its production is 
relatively well understood (see also Table 1), and this section provides a summary 
of results (\S 5.1) and a discussion of their implications 
(\S 5.2). 

\subsection{Summary of Results} 

We have estimated the enrichment levels of $^{40}$K from two different
scenarios. In the first case, circumstellar disks are enriched
directly by capturing ejecta from supernova explosions that detonate
within the same clusters, and a range of possible distributions for the
clusters are considered (see \S 3). For the most likely cluster distribution 
(a power-law distribution that extends up to stellar membership size
$N$ = $10^6$), we find modest enrichment levels. Only about 1 in 200
solar systems are predicted to double the abundance of $^{40}$K,
whereas 1 in 30 systems should receive a 10\% enhancement over the 
galactic background level.  The typical enrichment
levels of $^{40}$K fall in the range 0.01 to 1 p$M_\sun$, with the 
tail of the distribution extending up to 100 p$M_\sun$ (see Figure 7).

In addition to direct enrichment, we have (briefly) considered two
types of distributed enrichment. In one case, supernovae provide
additional $^{40}$K to the protostellar cores in a neighboring cluster
(\S 4.1). Since the cores are extended, they can subtend larger
solid angles (compared to circumstellar disks experiencing direct
enrichment) in spite of their larger distances. This scenario is
thus somewhat more effective than the case of direct enrichment. 
For example, about 1 in 80 solar systems are predicted to double 
their abundance of $^{40}$K (see Table 3).

We have also considered the entire molecular cloud as a dynamical
system (\S 4.2) and estimated the expected levels of enrichment
of $^{40}$K as the cloud evolves according to a simple model. In the
absence of losses --- assuming all of the $^{40}$K produced by supernovae
are retained within the cloud --- the later generations of star
formation can be significantly enhanced in $^{40}$K. The final generation
could have radioactive abundances up to a factor of about 5 times that of
the background galaxy. Only a relatively small fraction of the stars
are produced at the end of the cloud's lifetime, however, so that most
solar systems would be enhanced by smaller factors of $\sim 2-3$. When losses 
are included, these enrichment levels are even lower.  Keep in mind that this global model is included
for comparison, but has larger uncertainties than the calculations of
direct enrichment (\S 3) or from neighoboring clusters (\S
4.1).

\subsection{Discussion} 

The results of this paper have two important implications. The first
is that the enrichment of long-lived radioactive nulcei (LLRs) is
usually dominated by distributed enrichment mechanisms (rather than by
direct enrichment within the birth clusters of forming solar systems).
This finding is in contrast to the case of short-lived radioactive
nuclei, where direct enrichment and distributed enrichment can provide 
roughly comparable amounts of SLRs \citep{afnuke}. This result is not
unexpected, since the long half-lives of LLRs allow them to travel
much longer distances. 

The second implication of this work is that the fraction of solar
systems that experience substantial enrichment is of order one
percent. Specifically, this claim holds for $^{40}$K, which is one of
the most important nuclear species for planetary structure. In this
context, about one percent of solar systems receive enough LLRs to
double their abundance compared to the galactic background. Although
the neighboring cluster scenario is somewhat more effective (1 out of
80) than direct enrichment (1 out of 200), the results are roughly
comparable, and the former case contains more uncertainties. Of
course, doubling of the nuclear abundance only represents a useful
benchmark for comparison; the full description of radioactive
enrichment is provided by the distributions presented in \S 3
and \S4.

We note that these nuclear enrichment scenarios become more uncertain
as the distance from the supernovae (the source of LLRs) increases.
For solar systems within the same cluster as the supernova explosion,
we assume that the disks are efficient at capturing LLRs; in practice,
however, some losses will occur. Additional losses arise due to timing
issues, analogous to the case of SLR enrichment \citep{afnuke}. For
the neighboring cluster scenario, the abundances are calculated under
the assumption of efficient capture and mixing over the entire core;
additional calculations should explore the degree to which the
captured LLRs are delivered to the planet-forming disks at the end of
the star formation process. Some work along these lines has been
carried out for the case of SLRs (see, e.g.,
\citealt{desch,ouellette,boss} and references therein), but this
work should be generalized to the case of LLRs (where the time scales
and length scales are different).  Finally, distributed enrichment on
the scale of the entire molecular cloud has much larger uncertainties
than the other scenarios considered herein, and we have only
considered a simple model for comparison (see also Gounelle \& Meibom 2008,
Gounelle et al. 2009 for a related treatment of SLRs). To carry this work forward,
we need more realistic models of molecular cloud evolution, as well as 
better observational constraints on their lifetimes. 

Although only one percent of solar systems are predicted to experience
substantial LLR enrichment, the total number of highly enriched
systems in the Galaxy is quite large, of order $10^9$. Since
terrestrial planets are common, we expect a correspondingly large
number of them to also be enriched: The first Earth-sized planet in
the habitable zone of a main-sequence star has recently been
detected \citep{quintana} and projections suggest that about 10
percent of Sun-like stars harbor Earth-like planets in habitable
orbits \citep{petigura}. Favorably enriched planets would have ample
sources of internal heat, which helps to drive plate tectonics and
other geophysical activity. It is possible that such planets could
even be superhabitable, i.e., even more favorable for the development
of life than our own Earth \citep{heller}. This possibility should be 
kept in mind as we continue the search for habitable worlds. 

\noindent 
{\bf Acknowledgments:} This paper benefited from discussions with 
many colleagues including Konstantin Batygin, Juliette Becker, Ilse
Cleeves, Kate Coppess, Evan Grohs,  Minyuan Kay, and Dave Stevenson. 
We thank the referee for useful comments.  This work was
supported at the University of Michigan through the Michigan Center
for Theoretical Physics and at Xavier University through the Hauck
Foundation.

\end{document}